Research Article                                                    Open Access

## Morphogenesis of the Sternum in Quail Embryos

**Nashwa Araby[*1], Soha Soliman[1], Eman Abdel Raheem[2] and Yasser Ahmed[1]**

[1]Department of Histology, Faculty of Veterinary Medicine, South Valley University, Egypt,
[2]Department of Histology, Faculty of Medicine, South Valley University, Egypt

**Abstract**

The flat bone develops through intramembranous ossification, in which the mesenchymal cells are directly driven towards osteogenic lineage without the formation of cartilage template. While long bone develops through endochondral ossification, where cartilage template act as an intermediate stage between mesenchymal and bone tissues. Although the avian sternum is a flat bone, some studies describe formation of a cartilage template during its development. The aim of the current study was to observe the mechanism of ossification in quail sternum during embryonic development. Thirty quail embryos were collected for the current study (5 embryos/ day) during the period between Day (D) 5 and D10 of embryonic development and processed for light microscopy. The differentiation of mesenchymal condensation in to the chondrogenic cells was observed at D5 whereas the secretion of extracellular matrix could be evident at D6. The cartilage primordia were observed by D7 which were consisted of chondrocytes, embedded in matrix and surrounded by perichondrium. Later these primordia were developed in to cartilage template by D8 where the chondrocytes were present in their lacuna. This template attained the shape of future sternum by D9, which was more distinct at D10. These preliminary observations suggested that the quail sternum grows through endochondral ossification. The future study will further explore the histological changes of quail sternum during post-hatching development.

**Citation:** Araby *et al.,* 2018. Morphogenesis of the Sternum in Quail Embryos. SVU-IJVS, 1 (1): 16-24.

**Competing interest:** The authors have declared that no competing interest exists.

## Introduction

Development of flat and long bones occurs through intramembranous or endochondral ossification, respectively. During intramembranous ossification, the undifferentiated mesenchymal cells develop directly into bone tissue, while in the endochondral ossification, the mesenchymal tissue transforms into a cartilage template, on which the bone tissue is formed (Mescher, 2015). In the endochondral mode of ossification, a primary ossification center (POC) begins at the midshaft of the long bone, while a secondary ossification center (SOC) develops at the epiphysis (Ahmed and Soliman, 2013, Rivas and Shapiro, 2002). A cartilage growth plate develops close to the POC (Blair et al., 2002). The chondrocytes at the growth plate are organized in three zones; resting, proliferative, and hypertrophic (Blair et al., 2002, Mackie et al., 2008). In contrast to the endochondral ossification, the intramembranous ossification does not show POC or SOC and the bone is formed through direct transformation of undifferentiated mesenchymal tissue in to bone tissue (Little et al., 2011). Although the sternum is considered as an example of the flat bone (Martini et al., 2007), previous researches documented the formation of a cartilaginous model during its development (Nakane and Tsudzuki, 1999, Standring, 2016). The information regarding the histogenesis of sternum in quail seems to be meager thus, the current study aimed to explore the histological sequences of the quail sternum ossification during embronic development.

## Materials and methods

### Egg incubation

Fertilized quail (*Coturnix Japonica*) eggs were obtained from the production unit of quail, Faculty of Science, South Valley University, Qena, Egypt. The eggs were incubated at 37.5C° with a relative humidity 65 %. The eggs were rotated automatically every 6 hours from day 3 to day 10 of incubation.

### Sample processing for light microscopy

Eggs were carefully opened at the broad end and 5 embryos were daily collected at different stages of embryonic development (day 5, 6, 7, 8, 9 and 10). Samples were rinsed in phosphate-buffered saline (PBS), and fixed in 4 % neutral-buffered formalin for 24 hours, then in Bouin's fluid for 4-6 hours at 4C°. The embryo (as whole) were dehydrated gradually in ethanol, cleared in xylene the clearing agent and embedded in paraffin wax. Longitudinal and cross-serial sections (3-5µm thickness) were cut and stained with hematoxylin and eosin (H&E) according to (Bancroft et al., 1990).

## Results

The light microscopy revealed that the embryonic development of sternum started with mesenchymal cells condensation at the D5 of incubation. The mesenchymal cells were small and had many cytoplasmic processes (Fig. 1A-D). Later these cells were differentiated into chondrogenic cells or "prechondrocytes" by the D6 (Fig. 1E-H). The chondrogenic cells were semi-round or oval cells with a round or oval nucleus. The appearance of the cartilage primordia could be witnessed by D7 of incubation, which was consisted of chondrocytes located inside the lacuna and embedded in a distinguishable cartilage matrix and surrounded by an ill developed perichondrium (Fig. 2A-C). On D8, cartilage template was first appeared in which chondrocytes of central region of this cartilage template were large and round or oval in shape, furthermore, the perichondrium was distinct. While, prospective extremities contained differentiating chondrogenic cells with variable morphology. Some of them were flattened in shape surrounded by a small amount of cartilage matrix (Fig. 2D- F). By the D9, the cartilage template acquired the characteristic morphological features of the sternum. The rostral sterni and keel could be distinguished (Fig. 3A-C). The shape of the sternum became more prominent by the D10 of embryonic development. The cartilage template was very long and chondrocytes at the center were very distinct inside their lacuna and surrounded by abundant extracellular matrix. The perichondrium was well-developed except at the peripheral regions, where differentiating chondrogenic cells secreted a small amount of cartilage matrix (Fig. 3D-G). The histological changes of the quail sternum during embryonic development were summarized in (Fig. 4) and table (1).

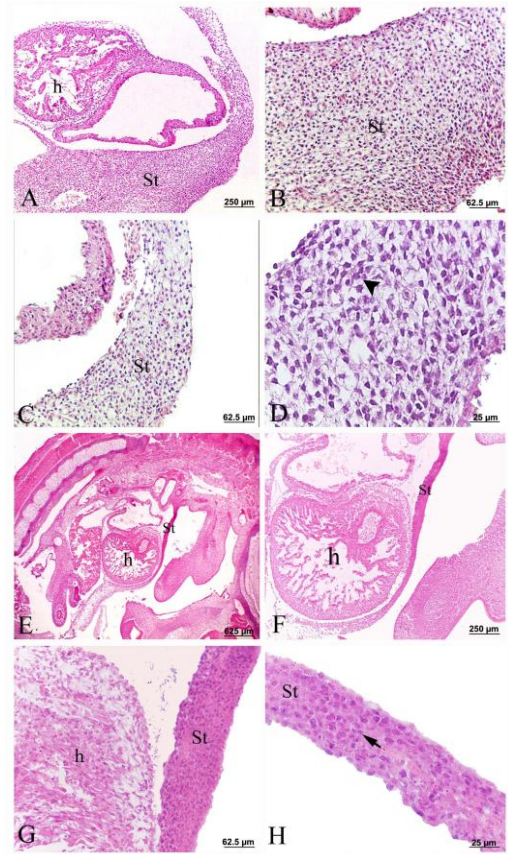

**Fig. 1**. Mesenchymal cells condensation and transformation into chondrogenic cells.

Paraffin sections from 5- (A- D) and 6- (E-H) day-quail embryos stained with H&E. Arrowhead in (D) indicates mesenchymal cell condensation, "St" refers to the place of developing sternum in embryo, "h" indicates Heart, arrow in (H) indicates chondrogenic cells secreted small amount of extracellular matrix. Bars = 250 μm in (A, F), 62.5 μm in (B, C and G), 25 μm in (D, H) and 625 μm in (E).

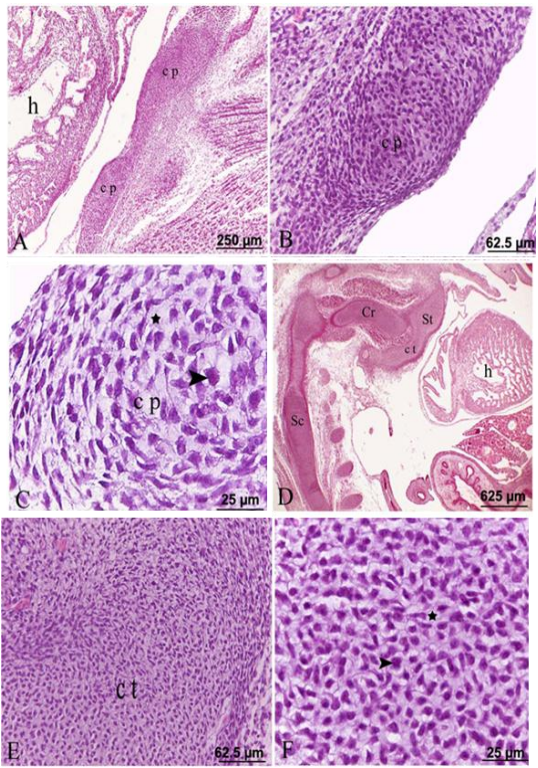

**Fig. 2.** Formation of the cartilage primordia and the cartilage template.

Paraffin sections from 7- (A-C) and 8- day (D-F) day-quail embryos stained with H&E. Note "cp" refers to cartilage primordia in (A-C) " h" in (A and D) refers to Heart, "St" in (D) refers to the developing sternum, "Cr" in (D) refers to developing crocoid bone, "Sc" in (D) refers to the developing scapula, "ct" in (D and E) refers to the cartilage template, arrowheads in (C and F) refer to the chondrocytes and the stars in (C and F) refer to extracellular matrix. Bars = 250 µm in (A), 62.5 µm in (B, E), 25 µm in (C, F) and 625 µm in (D).

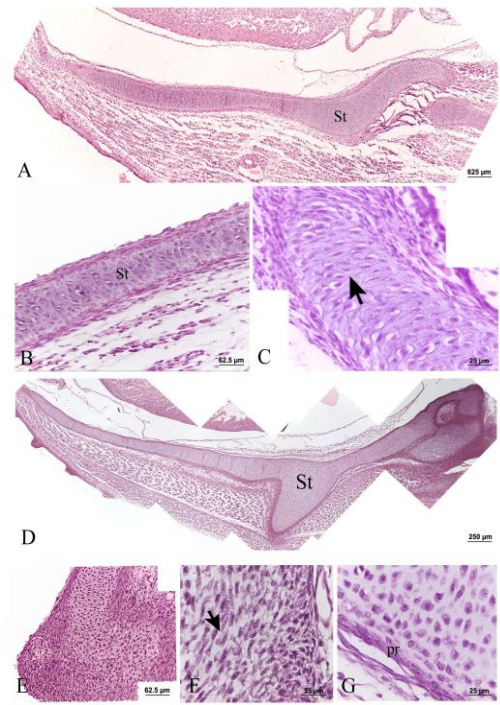

**Fig. 3.** Elongation of cartilage template and distinction the shape of the developing sternum.

Paraffin sections from 9-day (A-C) and 10-day (D-G) of embryonic development stained with H&E. Note "St" in (A, B and D) refers to developing sternum, arrows in (C and F) refer to prospective extremities of sternum and "pr" in (G) refers to the perichondrium. Bars = 625 µm in (A), 62.5 µm in (B, E), 25 µm in (C, F and G) and 250 µm in (D).

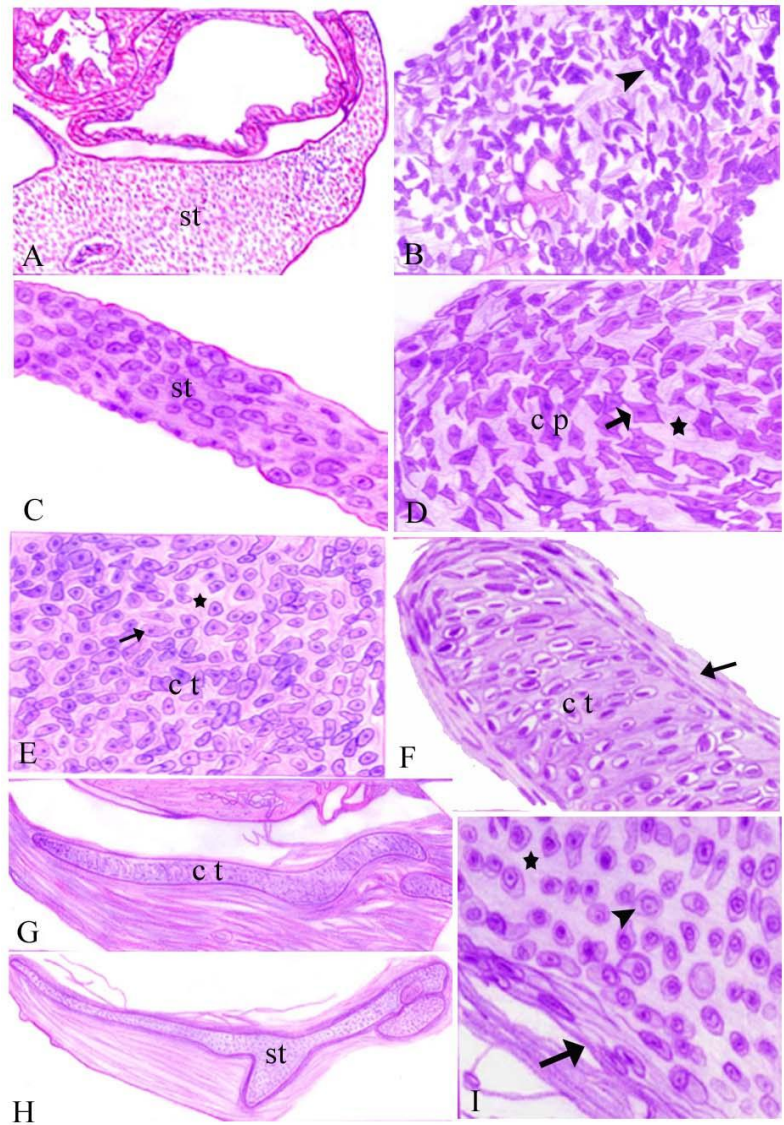

**Fig. 4.** Summary drawing for the stages of sternum development in 5-10-day-old quail embryos.

Mesenchymal cells condensation and its transformation into chondrogenic cells (A-C), "st" in (A) is prospective sternum. Arrowhead in (B) refers to mesenchymal cells condensation. Transformation into chondrogenic cells (C). Cartilage primordia (cp) in (D), arrow refers to a prechondrocyte and the star refers to cartilage matrix. Cartilage template (ct) is represented in (E-G), the prospective extremity of cells and matrix is illustrated in (F), arrow refers to a chondrocyte, the star refers to cartilage matrix in (E) and arrow in (F) refers to the perichondrium. The Distinction of the sternum shape is represented in (H and I); arrowhead refers to a chondrocyte inside the lacuna, the star refers to the cartilage matrix and arrow refer to the well-developed perichondrium in (I).

Table (1) Histological events of embryonic quail sternum development

| Embryonic developmental stages | |
|---|---|
| Developmental stage | Main histological events |
| Day 5 | • Mesenchymal cells condensation. |
| Day 6 | • Chondrogenic cells formation and secretion of a small amount of extracellular matrix. |
| Day 7 | • Appearance of cartilage primordial.<br>• Formation of chondrocyte lacunae<br>• The increase of matrix production.<br>• Formation of an ill-developed perichondrium. |
| Day 8 | • Development of the cartilage template.<br>• The increase of chondrocyte size and lacunae.<br>• Development of prospective extremities of cells and matrix. |
| Day 9 | • Cartilage template continued to grow and acquired the shape of the sternum. Spina external and crista sterni (carina or keel) have been more distinct. |
| Day 10 | • Distinction of the sternum shape.<br>• Further development of the cartilage template. |

**Discussion**

The aim of the current study was to explore the mode of ossification and to describe the stages of quail sternum ossification during embryonic life between D5 to D10 with the light microscopy. The current study reported 4 main stages of development; mesenchymal cells condensation, conversion into chondrogenic cells, appearance of cartilage primordia and formation of cartilage template. The manner of developmental sequences in sternum of quails was similar to the developmental sequences occurred in long bones via endochondral ossification as explained by (Miralles-Flores and Delgado-Baeza, 1990, Jeffcott and Henson, 1998, Gartner and Hiatt, 2001, Adams and Shapiro, 2002, Blair et al., 2002, Rivas and Shapiro, 2002, Mackie et al., 2008, Ahmed et al., 2015). In contrast to some literatures (Martini et al., 2007), where sternum was as an example of flat bone and developed by intramembranous ossification, others (Gilbert, 1997, Colnot et al., 2004, Little et al., 2011) mentioned that sternum is formed by fusion of two cartilaginous sternal plates which then ossified by advanced age in quails (Nakane and Tsudzuki, 1999) and in human (Standring, 2016). The current study revealed that sternum appeared firstly as mesenchymal cells condensation at D5 of incubation. Mesenchymal cells differentiated into chondrogenic cells at D6 of incubation and Cartilage primordia appeared after 7 days of incubation. Present findings were in an agreement with other study performed in quail embryo (Nakane and Tsudzuki, 1999) where, they identified sternal rudiments as a blue coloration after 6 days of incubation by alcian blue and alizarin red staining, which were used as special stains for cartilage and ossified bones, respectively. The sequence of developmental events of the sternum in quail embryos were similar to long bone, particularly, femur and tibia but differ in the time points of development (Ahmed et al., 2015, Ahmed and Soliman, 2013). For example, the condensed mesenchyme of the prospective femur and tibia appears at D5 and chondrification process occurs after 5.5 days. Cartilage primordia is formed at the D6 and cartilage templates are completely developed at D7 of the embryonic stage (Ahmed and Soliman, 2013). In the rabbit, long bone development is started by limb bud formation at D12 and cartilage primordia of the humerus appears by D14 of embryonic development. The cartilaginous diaphysis of the prospective humerus is observed by the D15. By the D16, the chondrocytes were clearly organized into proliferative and hypertrophic zones within the diaphysis (Ahmed et al., 2015). Chondrogenesis in chick embryo was reported to begin at D7.5-8 of incubation (Holder, 1978). While, in the mouse it appears at the D12.5 of embryonic development (Martin, 1990). In the current study, the formation of cartilage template, in which chondrocytes present inside lacuna was observed at D8 of incubation and continue to grow and take the shape of the characteristic shape of quail sternum between D9 and D10 of incubation. In the femur and tibia of quail embryos, cartilage template enlarged to form the shape of the future bone at the D6 of incubation (Ahmed and Soliman, 2013). A time difference is mentioned in the appearance of the ossified parts occurred in embryonic quail humerus at the 6th day of incubation and also could be measured using a double coloration technique (Pourlis et al., 1998).

In conclusion, the quail sternum is developed through endochondral ossification and not by intramembranous ossification as described for other flat bones. Further studies are required to follow the morphological changes of the sternum in late embryonic and post hatching development. These results should be considered during studies of gene expression associated with of quail development.